\documentclass{article}
\usepackage{amssymb,cite,multirow} 
\def\1ad{\mbox{\normalsize $^1$}}
\def\2ad{\mbox{\normalsize $^2$}}
\def\3ad{\mbox{\normalsize $^3$}}
\def\4ad{\mbox{\normalsize $^4$}}
\def\5ad{\mbox{\normalsize $^5$}}
\def\6ad{\mbox{\normalsize $^6$}}
\def\7ad{\mbox{\normalsize $^7$}}
\def\8ad{\mbox{\normalsize $^8$}}

\makeatletter
\@addtoreset{equation}{section}
\makeatother

\def\beq{\begin{equation}}                     %
\def\eeq{\end{equation}}                       %
\def\bea{\begin{eqnarray}}                     
\def\eea{\end{eqnarray}}                       
\def\nn{\nonumber}

\def\0 {\nonumber}

\def\N{\mathcal N}

\begin{document}
\setcounter{page}{0}
\begin{titlepage}
\titlepage
\rightline{hep-th/0606257}
\rightline{SPhT-T06/068}
\rightline{Bicocca-FT-06-12}

\vskip 3cm
\centerline{{ \bf \Large Gravity duals to deformed SYM theories }}
\vskip 0.5cm
\centerline{{ \bf \Large and Generalized Complex Geometry }}
\vskip 1.5cm
\centerline{Ruben
Minasian$^{a}$, Michela Petrini$^{b}$
and Alberto Zaffaroni$^c$}
\begin{center}
$^a$Service de Physique Th\'eorique,                   
CEA/Saclay \\
91191 Gif-sur-Yvette Cedex, France  
\vskip .4cm
$^b$ LPTHE, Universit\'es Paris VI et VII, \\
4 place Jussieu, 75252 Paris, France
\vskip .4cm
$^c$ Universit\`a di Milano Bicocca and INFN, sezione di
Milano-Bicocca \\
piazza della Scienza 3, Milano 20126, Italy
\end{center}
\vskip 1.5cm  
\begin{abstract}

We analyze the supersymmetry conditions for a class of 
SU(2) structure backgrounds of Type IIB supergravity, corresponding to
a specific ansatz for the supersymmetry parameters.
These backgrounds are relevant for the AdS/CFT correspondence since they are
suitable to describe mass deformations or beta-deformations of four-dimensional 
superconformal gauge theories. 
Using Generalized Complex Geometry we show that these geometries 
are characterized by a closed nowhere-vanishing vector field and a 
modified fundamental form which is also closed. 
The vector field encodes the information about the superpotential and the 
type of deformation - mass or beta respectively.
We also show that the Pilch-Warner solution 
dual to a mass-deformation of $\N =4$ Super Yang-Mills and the Lunin-Maldacena
beta-deformation of the same background fall in our class of solutions.

\end{abstract}

\vfill
\begin{flushleft}
{\today}\\
\end{flushleft}
\end{titlepage}

\newpage
\large

\section{Introduction}

Supergravity solutions with non-zero fluxes play an important role in any
attempt to recover 4-dimensional physics from string theory, from string 
compactifications to the AdS/CFT correspondence. In the
presence of type II Ramond-Ramond and/or Neveu-Schwarz fluxes, the internal
six dimensional geometry back-reacts and is typically not Ricci-flat. 
In the last few years many attempts have been done to find a geometrical
characterization of the internal manifolds analogue to the well-known
Calabi-Yau condition in the absence of fluxes. The formalisms of G-structures 
and Generalised Complex Geometry have led to some progress in this direction.
When applied to the AdS/CFT correspondence, the formalism of G-structures leads
to a classification of the geometrical structure of known solutions but,
more interestingly, can also be used to find new solutions.


The most elegant and best studied case is that of the structure group $G$
being SU(3). In this case it is possible
to give a set of general conditions in order for the solutions to be 
${\cal N}=1$ supersymmetric and a full classification of these backgrounds 
is known: in type IIB the manifold has to be complex, while
in type IIA it has to be twisted symplectic \cite{gmpt1}.
The geometry is fully characterized by a real
two-form and a complex three-form. One of the two invariant forms (the
two-form in IIA  and the three-form in IIB) is conformally closed, while
the non-closure of the other cancels against the fluxes.
The use of SU(3) structure allows to formulate the conditions for ${\cal
N}=1$ supersymmetry in a way \cite{gmpt1,gmpt2} which makes
the task of finding explicit solutions easier. And this turned out to be
particularly useful in the context of AdS/CFT correspondence. In
\cite{bgmpz} a family of regular SU(3)-structure equations was found
describing the baryonic branch of the Klebanov-Strassler solution
\cite{KS}.

Less can be said about the case of SU(2) structure solutions (which in
addition to the above mentioned forms admit a nowhere-vanishing vector
field) where the  considerable number of representations in the torsions
and fluxes makes the analysis using $G$-structures less powerful. In
particular, it has been shown that IIB backgrounds with SU(2) structure
are no longer required to be complex in order to preserve supersymmetry
\cite{dall'agata}. In the context of the AdS/CFT correspondence various
important solutions are characterized by SU(2) structures. In fact,
while conformal backgrounds of the form $AdS_5\times H$, with $H$ a 
Sasaki-Einstein manifold, and the corresponding non conformal backgrounds obtained by adding
fractional branes are described by SU(3) structures, massive and marginal
deformations of these conformal theories are typically characterized by
SU(2) structures and their geometry is still poorly understood.
It is one of the purposes of this paper to start
a detailed analysis of the conditions of supersymmetry related to
SU(2) structures. Having in mind applications to the AdS/CFT correspondence,
we will consider the case of type IIB solutions with non compact
internal manifolds. However, the geometrical characterization of
the SU(2) backgrounds described in this paper also have applications to the
compact case. 
  

We will make use of the language of Generalised Complex Geometry (GCG), which
is a  convenient conceptual framework for describing the ${\cal N}=1$
geometries. The basic objects here are {\sl pure spinors}, formal sums of even or odd forms, whose existence imposes certain topological conditions on the sum of the tangent and cotangent bundles of
 the internal manifold. In this language preservation of ${\cal N}=1$ supersymmetry reduces to a pair of differential conditions on the pure spinors which are somewhat schematically:
\bea
&& d_H \Phi_- = 0 \, ,\\
&& d_H \Phi_+ = *F \, .
\eea
Here $\Phi_+$ ($\Phi_-$) is the even (odd) pure spinor and $F$ is the formal sum of all RR fluxes. We will give their explicit expressions in Section 2. It is important that the pure spinors satisfy algebraic compatibility conditions in order to define a Riemannian metric on the internal space. Manifolds admitting a closed pure spinor (which as we see is one of the 
necessary conditions for ${\cal N}=1$ supersymmetry) are called Generalized Calabi-Yau (GCY) manifolds.


In this paper we discuss the supersymmetry equations (and Bianchi identities) 
for a particularly interesting class of SU(2)-structure backgrounds. We will
show that with a particular ansatz for the supersymmetry spinors the 
conditions for ${\cal N}=1$ supersymmetry give a very simple set of
equations for the SU(2) invariant forms and the fluxes. 
As all SU(2) structure backgrounds, these solutions are characterized by the 
existence of a conformally closed
vector. In addition to that, even if the metric is not K\"ahler, it is possible to
define 
a modified (1,1)-form which 
is conformally closed. We will show that two well known solutions in the
context of the AdS/CFT correspondence, the Pilch-Warner solution (PW) \cite{PW}, 
describing
a massive deformation of ${\cal N}=4$ SYM with an IR fixed point, and the
Lunin-Maldacena (LM) \cite{LM}, describing the marginal $\beta$-deformation of 
${\cal N} =4$ SYM, belong to this class of backgrounds. 
More generally, we expect to be able to characterize in terms of these 
special SU(2) structure  all 
other warped $AdS_5$ solutions with fluxes corresponding to (IR limits of)
massive deformations and $\beta$-deformations of conformal field theories
\footnote{A general set of equations for $AdS_5$ solutions in 
type IIB supergravity have been written in \cite{english}.}.
In this paper we will be mostly concerned with $AdS_5$ solutions, but 
our equations have applications to non conformal solutions as well. One
particularly important case of non conformal 
backgrounds which should belong to the
special class of solutions considered in this paper is given by the
general massive deformations of ${\cal N}=4$ SYM, ${\cal N}=1^*$. 

We will also study in detail the introduction of D3-brane probes in
the GCY backgrounds. The analysis of the supersymmetry conditions for a probe 
determines a (possibly empty) sub-variety of the internal manifold that is
in correspondence with the (mesonic) moduli space of vacua of the dual
gauge theory. We will show that all supergravity solutions dual
to a gauge theory with a non trivial moduli space of vacua necessarily
belong to the special class of SU(2) structures considered here.
  

The paper is organized as follows. In Section 2, we review the geometrical
characterization of SU(2) structure backgrounds and we introduce the general
tools that will be used in the rest of the paper. In Section 3 we specialize
to a particular spinorial ansatz and we write a very simple set of equations
involving the SU(2) invariant forms and the fluxes. In Section 4 we discuss
the geometry of the PW flow and the massive deformations of conformal
theories. We generalize a very simple class of type IIB supersymmetric
solutions found in \cite{GW}: a class of complex manifolds and associated fluxes which solve the
supersymmetry conditions of type IIB and are completely
characterized by the existence of a generalized K\"ahler potential.
These backgrounds can be used to describe massive flows to IR fixed points.
In this context we 
give some general conditions specifying $AdS_5$ solutions.
In section 5 we discuss the geometry
of the LM solutions and of (marginal) $\beta$-deformations of conformal
theories. We show that quite generally the action of T-duality on a
Calabi-Yau background leads to an SU(2) structure of the special case
considered in this paper.
Finally,  Appendix A contains the set of supersymmetry conditions for the most
general SU(2) structure ansatz  and Appendix B collects the formulae for the
T-duality which are used for the LM solution.

\section{SU(2) structure backgrounds}
\label{SU2struct}

In this paper we are interested in solutions of IIB supergravity of warped 
type
\beq
\label{metric}
\mbox{d}s^2 = e^{2 A} \mbox{d}s_4^2 + \mbox{d}s_6^2  \, ,
\eeq
where the internal manifold has SU(2) structure. 

An SU(2) structure manifold is characterized by the existence of 
two globally defined never-vanishing spinors which are never parallel
\beq
\label{spinorsSU2}
\eta_+  \quad  \quad \chi_+ = \frac{1}{2} z \cdot \eta_-  \, , 
\eeq
where $\eta_-$ is the complex conjugate of $\eta_+$ and $z \cdot$ denotes 
the Clifford multiplication by the one-form $z_m \gamma^m$.

Consequently the ten-dimensional supersymmetry parameters can be written as
\beq 
\label{spinors10d}
\begin{array}{c}
\epsilon_1 = \zeta_{+} \otimes \eta^1_{+} + \zeta_{-} \otimes \eta^1_{-} \, ,\\
\epsilon_2 = \zeta_{+} \otimes \eta^2_{+}
+ \zeta_{-} \otimes \eta^2_{-} \, ,
\end{array} 
\eeq
where $\zeta_{\pm}$ is a 4$d$ chiral spinor
 ($\zeta_{+}^{*} =  \zeta_{-}$) and the 6$d$ chiral spinors 
$\eta^{(i)}_{\pm}$ are related to the SU(2) structure spinors by
\beq
\label{spinoransatz}
\begin{array}{c}
\eta_{1 \, +} = a \eta_+ + b \chi_+ \, ,\\
\eta_{2 \, +} = x \eta_+ + y \chi_+ \, ,
\end{array}
\eeq
with $a,b,x$ and $y$ complex functions on the internal manifold. For the familiar SU(3) 
structure case corresponding to a Calabi-Yau manifold one has $x= - i a$ and $b=y=0$.
 
An alternative definition of SU(2) structure which will be useful in 
the following is given in terms of globally defined never-vanishing 
bilinears in the spinors (\ref{spinorsSU2})
\bea 
\label{formsSU2}
&& z =  -2 \chi_-^{\dagger} \gamma_m \eta_+ {\rm d}x^m \, ,\\
&& j =  \frac{i}{2} \chi^{\dagger}_+ \gamma_{mn} \chi_+ {\rm d}x^m \wedge 
{\rm d}x^n -  \frac{i}{2} \eta^{\dagger}_+ \gamma_{mn} \eta_+ {\rm d}x^m \wedge 
{\rm d}x^n \, ,\\
&& \omega =  - i \chi^{\dagger}_+ \gamma_{mn} \eta_+ 
{\rm d}x^m \wedge {\rm d}x^n  \, .
\eea
Here $z$ is a complex 1-form, $j$ and $\omega$ a real 2-form and a 
complex (2,0)-form satisfying
\bea
&& \omega \wedge j =0 \, ,\\
&& j^2 = \frac{1}{2} \omega \wedge \bar{\omega} \, , \\
&& z \llcorner j = z \llcorner \omega = 0  \, .
\eea

Each of the spinors $\eta_+$ and $\chi_+$ 
defines an almost complex structure compatible with the metric. The
associated (1,1)-forms are given by
\bea
J & = &  -  \frac{i}{2} \eta^{\dagger}_+ \gamma_{mn} \eta_+ {\rm d}x^m \wedge 
{\rm d}x^n \, =\, + j + \frac{i}{2} z\wedge \bar z \, ,\nonumber\\
\tilde J & = & - \frac{i}{2} \chi^{\dagger}_+ \gamma_{mn} \chi_+ {\rm d}x^m \wedge 
{\rm d}x^n \,  =\,  - j + \frac{i}{2} z\wedge \bar z \, ,
\eea
respectively. More generally, the SU(2) structure determines an entire 
U(1) family of almost complex structures compatible with the metric.
The corresponding (1,1)-forms are constructed in terms of the normalized
spinor 
\beq\label{red} \xi =\cos\delta \eta +\sin\delta \chi \eeq
as $J_{\xi}= -(i/2) \xi^{\dagger}_+ \gamma_{mn} \xi_+ {\rm d}x^m \wedge 
{\rm d}x^n$.

As shown in \cite{gmpt2}, by tensoring the supersymmetry parameters on the
internal manifold, $\eta^{1,2}_{\pm}$, it is possible to define formal sums
of even and odd forms respectively
\bea
\label{puresp}
&& \Phi_+ =  \eta^1_+\otimes \eta_{+}^{2 \, \dagger} \, ,\\
&& \Phi_- =  \eta^1_+\otimes \eta_{-}^{2 \, \dagger}
\eea
which are interpreted as pure spinors of $Cliff(6,6)$ in the context 
of Generalised Complex Geometry \cite{Hitchin,Gualtieri}.

For the choice of $\eta^1$ and $\eta^2$ in (\ref{spinoransatz}), the explicit form of the pure spinors reads
\bea
\label{purespSU2}
&& \Phi_+  = \frac{1}{8} \Big[ a \bar x e^{- i j} +  b \bar y e^{i j} - i
( a \bar{y} \omega + \bar{x} b \bar{\omega} ) \Big] e^{z \bar{z}/2} \, , \\
&& \Phi_{-}  = \frac{1}{8} \Big[ i (b y \bar{\omega} - a x \omega) + ( b x  
e^{i j} - a y e^{- i j}) \Big] z\, .
\eea
Following  \cite{gmpt2}, one can check explicitly that both $\Phi_+$ and  $\Phi_-$ are annihilated by six combinations of 
gamma-matrices and thus are pure. Since they have three annihilators in common, they are also  compatible. We would like to 
point out that the ansatz for the spinors and consequently the form of the pure spinors are slightly different from those used in \cite{gmpt2}. 
The two choices are related by a 
rotation of SU(2) structure that sets $b$ to zero in (\ref{spinoransatz}).
The form (\ref{purespSU2}) seems to be more suitable to describe the type of spinor ansatz that
appears in the supergravity solutions dual to mass deformations that we want to analyze in 
this paper.
Notice also that, in this form, the limits where the two spinors $\eta_{1,2}$ are always 
parallel 
(SU(3) structure) and always orthogonal (SU(2) structure) are both smooth.

As already mentioned in the Introduction, the supersymmetry variations of the supergravity fermions can then be 
re-expressed  as two equations for the two pure spinors 
\bea
\label{purespeq}
&& e^{-2A+\phi}(d-H\wedge) (e^{2A-\phi}\Phi_-) = 0  \, ,\nonumber \\
&& e^{-2A+\phi}(d -H\wedge) (e^{2A-\phi}\Phi_+) =  dA\wedge\bar\Phi_+ 
+\frac{e^\phi}{16}
\Big( a_- F_{\mathrm{IIB}} -i a_+ *F_{\mathrm{IIB}}\Big)\, ,
\eea
with $F_{\mathrm{IIB}}=F_1+F_3+F_5$ and 
$a_{\pm} = |a|^2 + |b|^2 \pm (|x|^2 + |y|^2)$. The functions 
$a,b,x,y$ are related to the norms of the pure spinors and satisfy
\beq
\label{norm}
\begin{array}{c}
\mbox{d} (|a|^2 + |b|^2) = (|x|^2 + |y|^2) \, \mbox{d}A \, ,\\
\mbox{d} (|x|^2 + |y|^2)=  (|a|^2 + |b|^2) \, \mbox{d}A \, .\\
\end{array}
\eeq
By expanding into forms of definite degree, a set of necessary conditions for 
${\cal N}=1$ solutions can be derived. The complete set of equations
corresponding to a generic ansatz (\ref{spinoransatz}) is reported in 
Appendix A.

The conditions for the existence of supersymmetric branes in a generalized
Calabi-Yau geometry have been studied in \cite{martucci}. A general requirement is that  the norms of the two spinors $\eta_{1,2}$ are equal. In our
notations, this translates into $a_-=0$, which, combined with (\ref{norm}), 
gives 
\beq
\label{general}
(|a|^2 + |b|^2) = (|x|^2 + |y|^2) = e^{A} \, .
\eeq
This condition has to be satisfied by all backgrounds arising as the near
horizon geometry of systems of branes.  

We will be interested in adding D-branes to the background, and, in particular,
space-time filling D3 branes. In the context of the AdS/CFT
solutions that we will consider, 
a supersymmetric D3 brane probes the (mesonic) moduli space 
of the dual gauge theory \footnote{It should be noted that, in general,  beside the mesonic branch there may exist
other vacua such as baryonic flat directions or
special non-abelian vacua related to a particular form  of the superpotential.
These however are not interpreted as D3 branes moving in the internal 
manifold.}. 
We are in fact interested in backgrounds 
which  originate from  a stack of $N$ D3 branes. The mesonic moduli space of
vacua is in correspondence with the supersymmetric distributions of $N$ branes
in different points of the internal manifold. In the familiar case of D3
branes probing a singular Calabi-Yau cone (with dual background $AdS_5\times
H$, where $H$ is the Sasaki-Einstein base of the cone) the moduli space is just
the symmetrized product of $N$ copies of the Calabi-Yau manifold. In more
general deformed backgrounds, fluxes can alter the supersymmetry conditions.
In particular they can introduce a superpotential for the probe that may
reduce the moduli space.
  
The supersymmetric conditions for space-time filling  Dp-brane
can be expressed in terms of the pure spinors (\ref{purespSU2}) as
\cite{martucci}
\bea
\label{Dp}
{\rm Im} ( i \Phi_{+}) \wedge e^{{\it F}-B}|_{\rm{top}} &=& 0 \, ,\nonumber\\
((dx^n + g^{nm} \iota_m) \Phi_{-}) \wedge  e^{F-B}|_{\rm{top}} &=& 0 \, ,
\eea
where ${\it F}$ is the world-volume gauge field. These equations, 
in the case of a D3 brane, become
\bea
\label{D3}
{\rm Im} ( i \Phi_{+})|_{(0)} &=& 0 \, ,\nonumber\\
\Phi_{-} |_{(1)} &=& 0 \, ,
\eea
where $\Phi_{\pm}|_{(k)}$ denotes the k-form component of the spinor. 
The two constraints can be interpreted as a D-term and an F-term conditions 
for the probe brane. For example, it has been shown \cite{martucci} that
the superpotential for the probe brane is given by
\beq
\label{supo}
{\mbox d} W = -i e^{2 A-\varphi} \Phi_-|_{(1)} = \frac{i}{8}e^{2 A-\varphi} (a y - b x) z \, .
\eeq

More explicitly, from equation (\ref{purespSU2}) we obtain the D-term and an F-term conditions 
for a supersymmetric D3 brane probe
\bea
\label{susyD}
{\rm Re} (a\bar x+b\bar y) &=& 0 \, ,\\
\label{susyF}
(b x -a y) &=& 0 \, .
\eea
The conditions for supersymmetry (\ref{purespeq}) imply
\footnote{Just take the real part of the one-form component of the
equation for $\Phi_{+}$ -- equation (\ref{1form}) in Appendix A.}
\beq
\label{dterm}
{\mbox d} (e^{A-\varphi} (a\bar x+b\bar y) ) = 0 \, .
\eeq
The quantity $(a\bar x+b\bar y)$ is therefore non-vanishing on the internal 
six-manifold. It follows that the D-term condition (\ref{susyD}) is
satisfied either everywhere or nowhere. In the cases where the condition 
(\ref{susyD}) is not satisfied, the moduli space is empty. When (\ref{susyD})
is satisfied, the F-term condition (\ref{susyF}) will select the sub-manifold
where a supersymmetric probe can freely move. This sub-manifold is in
correspondence with the mesonic moduli space of vacua of the dual gauge 
theory.

\section{A class of SU(2) structures}

In this paper we will focus on a specific form for the spinorial ansatz (\ref{spinoransatz}) which gives rise to a particularly simple set of supersymmetry
conditions. 

We will consider the ansatz
\bea
\label{spinoransatz2}
&& \eta_{1 \, +}  = a \eta_+ + b \chi_+ \, ,\nonumber\\
&& \eta_{2 \, +} =  -i (a \eta_+ - b \chi_+ ) \, ,
\eea
and, using equation (\ref{general}), we will parametrize
\bea
\label{spchoice}
&& a = i x = i e^{A/2} \cos \phi \, e^{ i \alpha} \, ,\\
&& b= -i y = -i e^{A/2} \sin \phi \, e^{ i \beta}  \, .
\eea

With this choice the supersymmetry conditions become very simple.
Some of the equations only contain the geometric data of the solution
\bea
\label{oneform}
&& \mbox{d}\Big(e^{3 A-\varphi} e^{i (\alpha + \beta)} \sin 2 \phi z \Big) =0 \, ,\\
\label{twoform}
&& \mbox{d} \Big[ e^{2 A -\varphi} \left ( j +\frac{i}{2} \cos2 \phi z \wedge \bar{z}\right ) \Big]=0  \, .
\eea
The other equations mix the geometry and the fluxes
\bea
\label{ffive}
&& e^{- 4 A+ \varphi} \mbox{d}\Big( e^{4 A- \varphi} \cos 2 \phi \Big) = - e^\varphi * F_5 \, ,\\
\label{fthree}
&& e^{- 4 A + \varphi} \mbox{d}\Big( e^{4 A - \varphi} \sin 2 \phi \, \mbox{Im} \hat{\omega} \Big) =
\cos 2 \phi H -  e ^{\varphi} *F_3 \, ,\\
\label{threeform}
&& \Big[\mbox{d}\Big( \frac{\cos^2 \phi \hat{\omega} 
+ \sin^2 \phi \bar{\hat{\omega}}}{\sin 2 \phi} \Big) + iH \Big] \wedge z = 0 \, ,\\
\label{fourform}
&& e^{- 2 A +\varphi} \mbox{d}  \Big(e^{2 A - \varphi}  \sin 2\phi \, \mbox{Im} \hat{\omega} \wedge z \wedge \bar{z} \Big) +2 i   H \wedge (j +\frac{i}{2} \cos2 \phi z \wedge \bar{z}) =0 \, ,\\
\label{Ifourform}
&& e^{- 4 A +\varphi} \mbox{d}\Big[ e^{4 A -\varphi} \left (  \cos 2 \phi \,  j^2 + i j \wedge z \wedge \bar{z}\right ) \Big] +
2  \sin 2 \phi H \wedge \mbox{Im} \hat{\omega} = e ^{\varphi} *F_1 
\eea
where $\hat{\omega} = e^{i (\alpha - \beta)} \omega$.

The geometrical conditions (\ref{oneform},\ref{twoform}) give the following
description of the six-dimensional internal space. It is convenient to 
define the Weyl rescaled six-dimensional metric
\beq
\label{rescaledmet}
\mbox{d} s_6^2 = e^{-2 A + \varphi} \mbox{d} \tilde{s}_6^2 \\ 
\eeq
which is the one seen by a D3 brane probing the background. The rescaled
internal manifold is characterized by a conformally closed  vector $z$.
In the almost complex structure determined by $\eta_+$, the metric is not 
K\"ahler, since ${\mbox d}J\ne 0$. However the modified two-form
\beq
\hat J = J - i \sin^2 \phi z \bar z = j + \frac{i}{2} \cos 2\phi z \bar z
\eeq
is closed.  

We can characterize our class of solutions as corresponding to 
 the  ${\cal N}=1$ backgrounds with a nontrivial moduli space for D3-brane probes.
This can be seen as follows. Given 
an SU(2) solution corresponding to the spinor ansatz 
(\ref{spinoransatz}), we can still make a redefinition of $\eta_+$ as in
(\ref{red}). The only effect would be a redefinition of the almost complex
structure (alternatively a redefinition of $J$, or $j,\omega$) we choose
in order to write our equations. There exist choices where the equations
simplify. However, not every spinor ansatz (\ref{spinoransatz})
can be reduced to the form considered in this Section.
A generic spinor ansatz (\ref{spinoransatz}) can be brought to the form
(\ref{spinoransatz2}) by a redefinition of $\eta_+$ if and only if
\beq \label{condition}
{\mbox Re} (a \bar x + b \bar y) = 0 \, .
\eeq

Equation (\ref{condition}) 
is exactly the D-term condition for the existence of a moduli space
of vacua for probe D3 branes. We recall from the previous Section that,
for a generic supersymmetric background, the condition (\ref{condition}) is satisfied 
in all points of the internal manifold or nowhere. This means that the
class of SU(2) structures we have just discussed contains 
at least 
all  ${\cal N}=1$ backgrounds  admitting 
a non trivial mesonic moduli space of vacua.

We can also consider the explicit form of the moduli space of vacua
for this class of SU(2) backgrounds.
The F-term condition (\ref{susyF}) simplifies to $ a b =0$ or
\beq
\label{moduli}
\sin 2 \phi = 0 \, .
\eeq 
This condition selects a (possibly empty) sub-manifold of the internal
space corresponding to the  moduli space of supersymmetric vacua of the dual 
gauge theory.


\section{Massive deformations of conformal gauge theories}
\label{relevant}

An interesting class of SU(2) structure backgrounds is provided by the duals 
of massive deformations of conformal gauge theories.

The typical example is ${\cal N}=4$ Super Yang-Mills to which we  can add a 
general supersymmetric mass deformation 
\beq
\int d\theta^2 dx^4 m_{ij} \Phi_i \Phi_j \, .
\eeq
On the supergravity side the massive deformation corresponds 
to a non-zero value of the complex 3-form $G_3 = F_3 - i e^{- \varphi} H$. 
It is known that, by deforming
${\cal N}=4$ SYM with a mass term for a single adjoint, the theory flows to
a fixed point. This follows from a standard argument due to Leigh and Strassler
\cite{LS}. The deformed theory has a superpotential
\beq
g \Phi_3 [\Phi_1,\Phi_2] + m \Phi_3^2
\eeq
which, after integrating out the massive field, becomes $ -\frac{g^2}{4 m} [\Phi_1,\Phi_2]^2$.
The conditions of conformal invariance, corresponding to
the vanishing of the beta-function and the requirement for the superpotential to have dimension three, combined with the obvious SU(2) symmetry of the theory,
require
\beq  
\Delta_{\Phi_{1,2}} (g,m) = \frac{3}{4} \, .
\eeq
This equation for the parameters $g$ and $m$ determines a line of IR 
fixed points \footnote{Actually, if we break the SU(2) symmetry, we can
find a larger manifold of fixed points.}. 
The IR conformal field theory has a complex two dimensional
moduli space of vacua parametrized by $\Phi_{1,2}$. 
The supergravity solution  corresponding to this flow has been studied in 
\cite{PW} and it will be referred to as the PW flow. 
It was originally obtained using
five-dimensional gauged supergravity and then lifted to ten dimensions.
The PW solution is complex, with constant dilaton and non trivial profile
for the antisymmetric three-form fields of type IIB. It reduces in the IR
to a warped $AdS$ background with three-form fluxes. At the IR fixed point,
the constant dilaton parametrizes the line of conformal field theories. 

It has been shown in \cite{GW} that the supersymmetry parameter  $\epsilon = \epsilon_1 + i \epsilon_2$ satisfies a 
dielectric-like projection
\beq
\label{dielectricpr}
\epsilon = \cos \phi \Gamma_{0123} \,  \epsilon 
+ i \sin \phi \,   \Gamma_{0123xy} \epsilon^*  \, ,
\eeq
where $x,y$ are internal directions.
Supersymmetry parameters in this class can be reduced to the ansatz 
(\ref{spinoransatz2}). 
A solution to the projection (\ref{dielectricpr}) is indeed 
given by a modification of the supersymmetry 
parameter of the undeformed background, $\epsilon_0=\xi_+\otimes \eta^0_{+}$,
\beq
\label{dielectricsp}
\epsilon = \cos \phi \epsilon_0 + i \Gamma_{xy} \sin \phi \epsilon_0^* \, .
\eeq
Here $\eta^0$ determines the complex structure of the undeformed background. 
In the case of ${\cal N}=4$ SYM we are dealing with $\mathbb{C}^3$; more generally, we may consider a Calabi-Yau cone corresponding to an ${\cal N}=1$ superconformal gauge theory. In all cases, $\Gamma_{xy}(\eta^{0}_+)^*$ can be rewritten
as $(\bar{z}/2) \cdot  \eta^0_{+}$ for a suitable complex vector $z$. 
From equation (\ref{spinors10d}) it follows that 
\bea
\eta^1_+ &\sim& \cos \phi \eta^0_+  + \sin \phi \frac{z}{2} \cdot \eta^0_- \, ,  \nonumber \\ 
\eta^2_- &\sim&  -i (\cos \phi \eta^0_+ - \sin \phi \frac{z}{2} \cdot \eta^0_-) \, ,
\eea
corresponding to the ansatz (\ref{spinoransatz2}). Thus
the mass deformation in the field theory 
selects one complex direction in the internal space corresponding to the
1-form $z$.

The full flow from ${\cal N}=4$ to the LS
fixed point is then described by an SU(2) structure with a spinor ansatz of the
form  (\ref{spinoransatz2}). In the following we will analyze in detail
the structure of the PW flow and, more generally, of 
deformations with three-form $G$ induced by a massive field. 
We will then specialize to the case of 
$AdS_5$ backgrounds corresponding to IR fixed points.

The ansatz (\ref{spinoransatz}) should cover more general cases of massive
deformations dual to $G_3$ fields which lead to non conformal theories
in the IR. The dual backgrounds of massive ${\cal N}=4$ SYM
are typically given by D3 branes dielectrically
expanded into a D5 branes via Myers effect \cite{myers}. The supersymmetry 
parameter should satisfy a projection similar to (\ref{dielectricpr}), 
where the first term can be interpreted as 
the standard D3 brane projection and the second as a D5 brane wrapping the $x,y$ directions in the internal manifold.

\subsection{A family of solutions of the supersymmetric conditions}

In this Section, we review and generalize a class of supersymmetric 
solutions of the equations of motion of type IIB supergravity found by the 
group in USC \cite{GW}. This class of solutions is particularly
suited for the description of massive deformations of conformal theories.
The original solutions \cite{GW} were obtained for metrics with $U(1)^3$
isometries; here we will extend them to more general metrics so that they may
be applied also to deformations of conformal theories associated with non toric
Calabi-Yau manifolds. We describe complex solutions 
of the equations (\ref{oneform})-(\ref{Ifourform}) that have at least
one $U(1)$ isometry corresponding to the gauge theory R-symmetry and a constant
dilaton corresponding to an exactly marginal direction of the conformal theory.
The only non-zero fluxes are the RR  5-form and the complex 3-form $G_3$.

We will work with the rescaled
six-dimensional metric we introduced in the previous Section 
(here the dilaton is set to zero)
\beq
\mbox{d} s_6^2 = e^{-2 A} \mbox{d} \tilde{s}_6^2 \, , 
\eeq
so that all the quantities in eqs (\ref{oneform})-(\ref{Ifourform}) are 
defined with respect to the metric $\mbox{d} \tilde{s}_6^2$. The equations
for the geometrical data read
\bea
\label{oneformR}
&& \mbox{d}\Big(e^{2 A} e^{i (\alpha + \beta)} \sin 2 \phi z \Big) =0 \, ,\\
\label{twoformR}
&& \mbox{d} \Big[ j +\frac{i}{2} \cos2 \phi z \wedge \bar{z} \Big]=0 \, .
\eea

We have seen that there is a preferred complex direction in the 
internal manifold specified by the conformally closed vector $z$. 
It is then natural to assume a four times two-dimensional splitting of the internal manifold
\beq
\label{metansatz}
\mbox{d} s_6^2 = \eta_i A_{ij} \bar{\eta}_j +  a_3 \eta_3 \bar{\eta}_3 \, ,
\eeq 
where the vielbein $\eta_3$ is proportional to $z$ 
\beq
\label{zz}
z = \sqrt{a_3} \eta_3 \, ,
\eeq
and the matrix $A$ is hermitian
\beq
\label{A}
A = \left(\begin{array}{cc}
a_1 & a_0 \\
\bar{a}_0 & a_2 
\end{array} \right) \, .
\eeq 
For the vielbeins we choose the following ansatz
\beq
\label{vielbein}
\begin{array}{l}
\eta_1 = \mbox{d} z_1 + \alpha_1 \mbox{d} z_3 \, \\
\eta_2 = \mbox{d} z_2 + \alpha_2 \mbox{d} z_3 \, \\
\eta_3 = u\mbox{d} z_3 \, .
\end{array}
\eeq
In the above expression 
$z_i$ are local complex coordinates 
($ z_{1,2} = h_{1,2}  + i \phi_{1,2}$, $z_3= \ln u + i \phi_3$) 
and $\alpha_i$ are complex functions.

In terms of the above vielbeins the 2-forms defining the SU(2) structure 
can be written as \footnote{It is also possible to introduce another set of 
vielbeins that make the metric diagonal $X_1 = \sqrt{a_1} \eta_1  + \frac{a_0}{\sqrt{a_1} }\eta_2$, 
$X_2 = \frac{\sqrt{\det{A}}}{\sqrt{a_1} } \eta_2$ and 
$X_3 = \eta_3$. The defining two-forms become $j= \frac{i}{2} ( X_1 \wedge \bar{X}_1 + X_2 \wedge \bar{X}_2)$ and $\omega = i X_1 \wedge X_2$.}
\bea
\label{j}
j &=& \frac{i}{2} A_{ij} \eta_i \wedge \bar{\eta}_j \, ,\\
\label{omega}
\omega &=&  i \sqrt{\det A } \eta_1 \wedge \eta_2 \, .
\eea

The form of the $z$ vector is determined by the massive deformation. 
Suppose we are adding
 the superpotential 
$W=\Phi_3^2$ for the adjoint field $\Phi_3$ to a conformal
field theory, for example ${\cal N}=4$ SYM. We identify the
complex coordinate $e^{z_3}=u e^{i\phi_3}$  with
$\Phi_3$ in the 
supergravity solution. From equation (\ref{supo}) we find
\beq 
e^{2 A} e^{i (\alpha + \beta)} \sin 2 \phi  z \sim {\mbox d} W = {\mbox d} ( u^2 e^{2 i\phi_3}) \, ,
\eeq
so that 
\bea
\label{conf}
&& e^{2 A} \sqrt{a_3} \sin 2 \phi = m u \, ,\nonumber\\
&& e^{i(\alpha+\beta)}=e^{2 i \phi_3} \, ,
\eea
where $m$ is a constant. 
This automatically solves equation (\ref{oneformR}). 

From eq. (\ref{twoformR}) it follows that there exists a closed 2-form
\bea
\label{jhat}
\hat{J} &=& j +\frac{i}{2} \cos2 \phi z \wedge \bar{z} \\
&=& \frac{i}{2} A_{i j} \eta_i \wedge \bar{\eta}_j +
\frac{i}{2} \cos2 \phi \, a_3 \eta_3 \wedge \bar{\eta}_3 \, . \nn
\eea
Therefore, although the metric is not K\"ahler, we can introduce, 
at least locally, a generalized K{\"a}hler potential $F$
\beq
\label{Jkal}
\hat{J} = \frac{i}{2} \frac{\partial^2 F}{\partial z_i \partial \bar{z}_j } 
\mbox{d} z_i \mbox{d} \bar{z}_j \, .
\eeq

Comparing (\ref{Jkal}) and (\ref{jhat}), it is possible to express the functions $a_i$ in
the metric ansatz in terms of the generalized K\"ahler potential
\bea
\label{ai1}
&& A_{ij} =  \frac{\partial^2 F}{\partial z_i \partial \bar{z}_j }  
\quad i,j=1,2 \, , \\
\label{ai2}
&& A_{ij} \bar{\alpha}_j =   \frac{\partial^2 F}{\partial z_i \partial \bar{z}_3 }  \, ,\\
\label{ai3}
&& \alpha_i A_{ij}  =    \frac{\partial^2 F}{\partial 
\bar{z}_i \partial z_3 } \, ,\\
\label{ai4}
&& u^2 a_3 \cos 2\phi + \alpha_i A_{ij} \bar{\alpha}_j =   
\frac{\partial^2 F}{\partial z_3 \partial \bar{z}_3 } \, .
\eea

It remains now to solve the equations involving the fluxes
\bea
\label{ffiveb}
&& e^{- 4 A} \mbox{d}\Big( e^{4 A} \cos 2 \phi \Big) = - * F_5 \, ,\\
\label{fthreeb}
&& e^{- 4 A} \mbox{d}\Big( e^{2 A} \sin 2 \phi \, \mbox{Im} \hat{\omega} \Big) =
\cos 2 \phi H - *F_3 \, ,\\
\label{threeformb}
&& \Big[\mbox{d}\Big( \frac{e^{- 2 A}}{\sin 2 \phi} (\cos^2 \phi \hat{\omega} 
+ \sin^2 \phi \bar{\hat{\omega}} ) \Big) + iH \Big] \wedge z = 0 \, ,\\
\label{fourformb}
&& d  \Big(e^{- 2A}  \sin 2\phi \, \mbox{Im} \hat{\omega} \wedge z \wedge \bar{z} \Big) +2 i   H \wedge (j +\frac{i}{2} \cos2 \phi z \wedge \bar{z}) =0 \, ,\\
\label{Ifourformb}
&& \frac{1}{2} e^{- 2 A} \mbox{d}\Big(  \cos 2 \phi \,  j^2 
+ i j \wedge   z \wedge \bar{z} \Big) +
 \sin 2 \phi H \wedge \mbox{Im} \hat{\omega} =0 \, .
\eea

The first of the equations involving fluxes, (\ref{ffiveb}), can be considered as 
a definition of the five-form flux. To solve the other ones 
we introduce an ansatz
for the complex three-form which is a straightforward generalization of that in \cite{GW}
\beq
\label{G3ansatz}
G_3 = \mbox{d} A_2 \, ,
\eeq
with $A_2 = C_2 - i B_2$ and 
\beq
\label{A2ansatz}
A_2 = \frac{2 i}{m} e^{z_1 + z_2 -\bar{z}_3} [
\mbox{d} z_1 \wedge  \mbox{d} z_2  -\sin^2 \phi \, \eta_1 \wedge \eta_2 ] \, .
\eeq

The above choice solves (\ref{threeformb}) with the condition
\beq
\label{phase}
e^{z_1 + z_2 -\bar{z}_3} = m  e^{ i ( \alpha - \beta)} 
\frac{\sqrt{\det A}}{e^{2 A}\sin 2\phi} \, ,
\eeq
or, expanding the coordinates in real and imaginary part $z_i=h_i+i \phi_i$,
\bea
\label{phaser}
e^{h_1+h_2} &=& m \sqrt{\det A a_3} \, ,\\
\label{phasei}
e^{i(\alpha-\beta)}& =&e^{i(\phi_1+\phi_2+\phi_3)} \, .
\eea
The meaning of this condition is that
\beq 
\label{Omega} 
\hat\omega \wedge z = \frac{i}{m}e^{z_1+z_2+z_3} \mbox{d}z_1 \wedge \mbox{d}z_2 \wedge \mbox{d}z_3 \, ,
\eeq
so that SU(3) structure holomorphic three-form $\Omega=\omega \wedge z$  is closed, up to a phase which can be reabsorbed in the definition of the vielbeins $\eta_i$.  

From (\ref{fourformb}) one obtains
\beq
\label{cond1}
\left\{ \begin{array}{l}
\frac{\partial}{\partial \bar{z}_i} ( a_3  u^{2} \sin^2 \phi ) = 
- \alpha_j A_{ji} \, ,\\
\frac{\partial}{\partial z_i} (a_3  u^{2}  \sin^2 \phi) 
= -  \bar{\alpha}_j \bar{A}_{ji} = - A_{ij} \bar{\alpha}_j \, ,
\end{array}
\right.
\eeq
which can be further simplified, using (\ref{ai2}) and (\ref{ai3}), to give
\beq
\label{cond1b}
\left\{ \begin{array}{l}
\frac{\partial}{\partial \bar{z}_i} ( a_3   u^{2}  \sin^2 \phi 
+ \frac{\partial}{\partial z_3} F) = 0 \, ,\\
\frac{\partial}{\partial z_i} (a_3   u^{2}  \sin^2 \phi 
+  \frac{\partial}{\partial \bar{z}_3} F ) = 0 \, .
\end{array}
\right.
\eeq
Finally (\ref{Ifourformb}) gives
\beq
\label{cond2}
\frac{\partial}{\partial z_3} (a_3 u^2 \sin^2 \phi 
+  \frac{\partial}{\partial \bar{z}_3} F ) = 0 \, .
\eeq

For backgrounds that do not depend on $\phi_3$ we can solve all the
previous equations by taking
\beq
\label{fl} 
a_3 u^2 \sin^2 \phi = -  \frac{\partial}{\partial z_3} F \, .
\eeq
At this point a long but straightforward computation shows that equation
(\ref{fthreeb}) is also satisfied.

Notice that the solution is completely specified by a generalized K\"ahler potential $F(z_i)$ that satisfies the condition of closure for $\Omega$, equation
(\ref{phaser}). 
Indeed, the metric is determined by $F$, the fluxes by equation 
(\ref{ffiveb}) and by the ansatz (\ref{G3ansatz}),
$\phi$ by condition (\ref{fl}), the
warp factor $A$ by equation (\ref{conf}) and, finally, 
 the two phases $\alpha$ and 
$\beta$ by (\ref{conf}-\ref{phasei}). The situation is similar to the
Calabi-Yau case where the solution is determined by a K\"ahler potential
which satisfies the condition of closure for $\Omega$. We see that, in
this particular class of solutions, the inclusion of fluxes does not
introduce new constraints.

The equations further simplify if the metric has three
$U(1)$ isometries. This is the case where we introduce a mass deformation
in ${\cal N}=4$ SYM or more generally in a theory dual to a toric Calabi-Yau.
The mass term typically breaks one global symmetry of the theory, but 
in the supergravity solution
this breaking  can be included in the phase of
the $G$ field and does not affect the metric. We can then have a solution 
where $F(h_i)$ does not depend on the angles $\phi_i$, the metric has a
$U(1)^3$ isometry and the $G$ field has a phase $e^{i(\phi_1+\phi_2+\phi_3)}$
according to equation (\ref{G3ansatz}). In this toric cases, $\hat J$ can
be rewritten as
\beq
\hat J = \frac{1}{4} \frac{\partial^2 F}{\partial h_i \partial h_j} \mbox{d} h_i \wedge \mbox{d} \phi_j = \mbox{d} \left ( \frac{1}{4} \frac{\partial F}{\partial h_i} \mbox{d} \phi_i \right ) \, .
\eeq
Here we used $\mbox{d}z_i=\mbox{d}h_i + i \mbox{d}\phi_i$, $h_3=\log u$.
The quantities $\frac{\partial F}{\partial h_i}$ then play the role of 
momentum map variables for the $U(1)^3$ fibration. 
The PW flow belongs to this toric class of solutions.

\subsection{$AdS_5$ solutions}

We obtain $AdS_5$ solutions by restricting to warped conical  
six-dimensional metrics,
\beq\label{cone} 
d\tilde s_6^2 = H^4 dr^2+r^2 ds_5^2 (y)
\eeq
where $H$ only depends on the angular coordinates $y_i$ of the base
$ds_5$. In this way the full 10-dimensional metric
\beq
ds_{10}^2 = e^{2 A} ds_4^2 + e^{-2 A} [H^4 dr^2+r^2 ds_5^2]
\eeq
factorizes to the warped AdS form
\beq\label{warpedAdS}
H^2 ds_{AdS}^2 + H^{-2} ds_5^2 \, ,
\eeq
provided that
\beq
\label{warping}
e^{2 A} = r^2 H^2 \, .
\eeq
Conformal invariance requires $F\sim r^2$, $A_{ij}\sim r^2$ 
and $\alpha_i$ adimensional.
The correct scaling behavior of the metric combined with equation (\ref{conf})
also requires $u\sim r^{3/2}$ and $a_3\sim 1/r$. This is a consequence
of the marginality of the superpotential term $\Phi_3^2$ at the fixed point, 
which in turn implies dimension $3/2$ for the field $\Phi_3$ \footnote{This is 
better thought as the composite $[\Phi_1,\Phi_2]$ at the IR fixed point.}
associated with the variable $u$. It is known that, for a K\"ahler cone,
$F=4 r^2$ \cite{MSY}. In our case, more generally,
\beq 
F=r^2 f(y) \, ,
\eeq
 where $f$ is a function on the five-dimensional base. The conditions for the
existence of an $AdS_5$ solution finally require the absence of terms
linear in $\mbox{d}r$ in the six-dimensional metric. This 
represents a further constraint on the coordinates $z_i$ and the K\"ahler
potential $F$.

All these conditions considerably simplify in the toric case. We can 
choose coordinates $\lambda,k,\phi_i$ on the five-dimensional base such that
\bea
\mbox{d} z_i &=& \mbox{d} h_i + i \mbox{d} \phi_i = n_i \frac{\mbox{d} r}{r} + \mbox{d} v_i(\lambda,k) + i \mbox{d} \phi_i\, , \qquad i=1,2 \nonumber\\
\mbox{d} z_3 & = & \frac{\mbox{d} u}{u} +i \mbox{d} \phi_3 = n_3 \frac{\mbox{d} r}{r} + \frac{\mbox{d}\lambda}{\lambda} + \mbox{d} \phi_3\, , \qquad n_3=\frac{3}{2} \, .
\eea
The generalized K\"ahler potential only depends on $h_i$, $F(h_i)$. We require
$n_1+n_2+n_3=3$ in order for the (3,0) holomorphic 
form $\Omega$ to scale as $r^3$ (cfr equation (\ref{Omega})).

Determining the metric from $J=\hat J +(1-\cos 2 \phi) i z\bar z/2$
\beq 
ds_6^2 = \frac{1}{4}\frac{\partial^2 F}{\partial h_i \partial h_j} dz_i  d\bar{z}_j +(1-\cos 2 \phi) a_3 u^2 dz_3  d\bar{z}_3 \, ,
\eeq
and using repeatedly the condition of scale invariance
\beq 
n_i \frac{\partial F}{\partial h_i} = r \frac{\partial F}{\partial r} = 2 F \, ,
\eeq
we obtain the condition for the absence of mixed terms in $dr$
\beq
\label{mixed}
\mbox{d} F= 2 F \frac{\mbox{d}r}{r} + 3 u \frac{\partial F}{\partial u} \frac{\mbox{d}\lambda}{\lambda} \, .
\eeq
This implies that
\beq F = r^2 f(\lambda) \eeq
and 
\beq 
\label{constraint}
\lambda f'(\lambda) = \frac{3}{r^2} u \frac{\partial F}{\partial u} \, .
\eeq
Finally, the warp factor is determined by the $dr^2$ term in the metric which,
with simple manipulations, can be written as
\beq 
H^4 = f(\lambda) -\frac{3}{4}\lambda f'(\lambda) \, .
\eeq

As an example we can reconstruct the PW solution for the IR fixed point
of massive ${\cal N}=4$ SYM. The undeformed solution is associated with
$\mathbb{C}^3$ for which we choose coordinates
\bea
&& e^{z_1}=r \cos\theta \cos \varphi e^{i\phi_1} \, , \nonumber\\
&& e^{z_2}=r \cos\theta \sin \varphi e^{i\phi_2}\, , \nonumber\\
&& e^{z_3}=r \sin\theta e^{i\phi_3} \, .
\eea
The complex coordinates for the IR fixed point are just obtained with
a rescaling
\bea
&& e^{z_1}=r^{3/4} \cos\theta \cos \varphi e^{i\phi_1} \, ,\nonumber\\
&& e^{z_2}=r^{3/4} \cos\theta \sin \varphi e^{i\phi_2} \, ,\nonumber\\
&& e^{z_3}=r^{3/2} \sin\theta e^{i\phi_3} \, .
\eea
We can choose $\lambda=\sin \theta$ and $k=\varphi$. The solution 
of the constraint (\ref{constraint}) is  obtained by taking
\beq 
F = c r^2 (1- 2 \sin^2 \theta ) \, ,
\eeq
from which $H^4=c (1+\sin^2\theta)$. All other quantities are
consistently determined by the equations in the previous Section. We obtain
for example
\beq
 \cos^2 \phi =\frac{\sin^2\theta}{2(1+\sin^2\theta)} \, .
\eeq
The full PW metric can be reconstructed as
\bea
 d\tilde s_6^2 &=& \frac{3}{4} dr^2 (1+\sin^2 \theta ) + r^2 \Big[ \frac{1}{2}(1+\sin^2 \theta )d\theta^2
+ \cos^2 \theta d\phi^2 + \cos^2 \theta (\cos^2 \phi d\phi_1^2 +\sin^2 \phi d\phi_2^2)  \nonumber\\
& + & \frac{\sin^2 \theta}{2} d\phi_3^2 + \frac{(\cos^2 \theta
(\cos^2 \phi d\phi_1 +\sin^2 \phi d\phi_2) + \sin^2 \theta d\phi_3)^2}{ 3 (1+\sin^2\theta)}\Big]
\eea 
which is equivalent to formula (7.8) of the second paper in \cite{GW}.

The moduli space of vacua of the dual gauge theory is two-dimensional and
spanned by the vacuum expectation values of the fields $\Phi_{1,2}$. The
moduli space for D3-brane probes of the PW solution is obtained by
imposing equation (\ref{moduli})
\beq
\sin 2 \phi = \frac{\sin\theta \sqrt{2 +\sin^2\theta}}{1+\sin^2\theta} \equiv 0 \, ,
\eeq
which selects the two-dimensional sub-manifold $u=r^{3/2}\sin\theta \equiv 0$.
On this sub-manifold $\cos 2 \phi\equiv 1$ and
\beq \hat J \equiv J \, ,\eeq
so that the moduli space seen by the probe is a two-dimensional K\"ahler
manifold as required by supersymmetry for ${\cal N}=1$ gauge 
theories \footnote{An explicit computation of the K\"ahler potential for a probe in
the PW flow was performed in \cite{johnson}.}.

It would be interesting to look for other warped $AdS_5$ solutions of type IIB
supergravity. The above analysis has been 
performed for massive deformations of ${\cal N}=4$ Super Yang-Mills,
thus for deformation of $AdS_5 \times S^5$.  
However it is likely to describe also mass deformations of ${\cal N}=1$
quiver gauge theories associated with more general Sasaki-Einstein manifolds $H$  and backgrounds of the form $AdS_5 \times H$. In quiver gauge theories,
the addition of a mass term
for a single adjoint field leads quite generically to an IR fixed point
if there are enough global abelian symmetries \footnote{The conditions for 
conformal invariance of the original theories are
determined only up to free parameters associated with the abelian symmetries,
since the latter can mix with the R-symmetry. The exact dimensions of
the chiral fields is determined using a-maximization \cite{amax}. Upon
the addition of a mass term, we should just restrict the maximization
on the sub-variety where the R-charge of the massive field is one. 
We thus expect a solution of the a-maximization if the original theory
have enough abelian symmetries.}. This is typically the case for
 Sasaki-Einstein backgrounds where massless vectors arise not only from isometries but also from RR fields. We thus expect a large number of $AdS_5$
solutions dual to massive deformations of conformal theories. Some of them
will be still described by Sasaki-Einstein backgrounds \footnote{
Not all the mass deformations are dual to modes of the three-form 
$G$ field in supergravity, as it happens for  ${\cal N}=4$ SYM. For a generic 
quiver theory, where $H$ may have orbifold singularities, 
some mass deformations are dual to geometrical blowing up modes. 
A familiar example is provided by the
${\cal N}=2$ quiver gauge theory associated with the singular Calabi-Yau
$\mathbb{C}^2/Z_2\times \mathbb{C}$, i.e. $H=S^5/Z_2$. A mass deformation
dual to the blow up mode leads to an IR fixed point corresponding to the
conifold.
In this cases, the IR fixed point will be still of the form $AdS_5\times H$
with $H$ a Sasaki-Einstein manifold.} but others will be described by
warped solutions with non zero $G$ flux. At the moment, the only explicitly
known example in the latter class is PW and it would be quite interesting to
find alternative ones. The number of known Sasaki-Einstein metrics has been
recently enlarged with the discovery of the $Y^{pq}$ and $L^{pqr}$ metrics 
\cite{YL}. In particular
the quivers associated with the Generalized Conifolds, which are particular 
cases of the $L^{pqr}$ \cite{YLgauge}, have adjoint fields and massive deformations leading to IR fixed points, which could correspond to warped $AdS_5$ solutions with fluxes. 

\section{Marginal deformations of conformal field theories}

A second interesting class of SU(2) structure backgrounds is provided by the
duals of marginal deformations of conformal gauge theories.

Once again we consider the example of ${\cal N}=4$ SYM. It is known that there exists a manifold of ${\cal N}=1$ 
fixed points that contains the
${\cal N}=4$  Yang-Mills theory \cite{LS}. 
The corresponding theories can be
described in  ${\cal N}=1$ language as containing the same fields as ${\cal N}=4$  SYM but with a
superpotential (modulo the SU(3) global symmetry)  
\beq
\label{beta2}
h \,\rm{Tr}( e^{ i  \pi \beta} \Phi_1\Phi_2 \Phi_3 - e^{- i  \pi \beta}
 \Phi_1\Phi_3 \Phi_2  ) + h' \rm{Tr}( \Phi_1^3 +\Phi_2^3 +\Phi_3^3 ) \, .
\eeq
In addition to the three parameters appearing in the superpotential, we have
the complexified coupling constant.
There is a particular relation between the four complex parameters
$g_{YM},h,h',\beta$ for which
the theory is superconformal \cite{LS}. 
The equations for the vanishing of all beta-functions are
satisfied if
\beq 
\Delta_{\Phi_i} (g_{YM},h,h',\beta)=1 \, .
\eeq
Using the obvious permutation symmetry among the $\Phi_i$ we conclude that
we have a single equation for four unknowns. This yields a
three-dimensional complex manifold of fixed points.

We consider here the case of the $\beta$-deformation with $h^\prime =0$,
\beq 
h \,\rm{Tr}( e^{ i  \pi \beta} \Phi_1\Phi_2 \Phi_3 - e^{- i  \pi \beta}
 \Phi_1\Phi_3 \Phi_2  ) \, ,
\label{beta}\eeq
whose supergravity dual has been found in \cite{LM} and it will be referred 
as the  LM solution. The theory still 
preserves a $U(1) \times U(1)$ global symmetry, in addition
to the R-symmetry. The global $U(1) \times U(1)$ acts with charges
$(0,1,-1)$ and $(-1,1,0)$ on the three chiral fields $\Phi_i$.
The deformation of the superpotential modifies the F-term equations
and reduces the moduli space of vacua of the theory.
The F-terms read
\begin{equation}
\Phi_1\Phi_2=e^{-2\pi i \beta}\Phi_2 \Phi_1, \,\qquad 
\Phi_3\Phi_1=e^{-2\pi i \beta}\Phi_1 \Phi_3,
\, \qquad \Phi_2\Phi_3=e^{-2\pi i \beta}\Phi_3 \Phi_2 \, .
\end{equation}
The original moduli space of $\N=4$ SYM was parametrized by arbitrary
diagonal matrices $\Phi_i$. In the $\beta$-deformed theory, we see 
that a diagonal matrix for, say, $\Phi_i$ solves the F-term equations
only if the other two $\Phi_k$ vanish. The mesonic moduli space is then the
union of three branches meeting at the origin. A D3-probe sees a 
three-dimensional moduli space isomorphic to $\mathbb{C}^3$ in
 the case of $\N=4$
SYM but only three complex lines intersecting at the origin in the case
of the $\beta$-deformed theory. It is known \cite{bl} that, for special values of the deformation parameter where $\beta$ is rational, we can have other Coulomb branch vacua where $\Phi_i$ define a non-commutative torus. As usual, we do not consider in this paper baryonic type vacua.

We will further restrict to the case where $\beta$ is real. The case where $\beta$ is complex can be obtained by a further type IIB 
S duality. 
The case of real $\beta$ is particularly interesting 
because it can be obtained from the ${\cal N}=4$ solution by a T-duality. 
The T-duality transformation acts on the two-torus made with two 
$U(1)\times U(1)$ isometries of the original solution. 
As we will show in the following, a T-duality on two angular directions 
corresponding to isometries of $\mathbb{C}^3$, or more generally of a
Calabi-Yau, transforms the original SU(3) structure into an SU(2) structure satisfying the special ansatz (\ref{spinoransatz2}), thus of the special form considered in this paper. 

\subsection{Spinors and the action of T-duality}

The generating solution technique used in \cite{LM} applies to all backgrounds
with at least two isometries. We call $\varphi_{1,2}$ 
the two corresponding angles and $g_{ij}, B_{ij}\equiv B\epsilon_{ij}\, (i,j=1,2)$ 
the metric and the antisymmetric NS two-form on the $T^2$ spanned by  $\varphi_{1,2}$. 
The metric thus reads 
\beq\label{met}
ds_6^2 = g_{ij} e_{\varphi_i} e_{\varphi_j} + d\tilde s_4^2 = y_1^2+y_2^2 + d\tilde s_4^2  \, ,
\eeq  
where the one-forms $e_{\varphi_i}=d\varphi_i+...$ have been used to eliminate off-diagonal terms, 
if needed, and $d\tilde s_4^2$ does not depend on $\varphi_i$. $y_{1,2}$ correspond to a choice of vielbeins along the $T^2$.

In order to obtain new solutions, we can use the SL(2,R) subgroup of the T-duality group O(2,2)
that acts on the complexified
K\"ahler modulus $\nu = B +i \sqrt{g}$ of the two torus as 
\beq \nu \rightarrow \frac{a \nu +b}{c \nu +d} \, .\eeq
As argued in \cite{LM}, the particular
element
\begin{equation}\label{mlsl2}
LM=\left(\begin{array}{cc}
1 & 0 \\
\gamma  & 1
\end{array}\right)
\end{equation}
transforms regular solutions in other completely regular solutions depending
on the parameter $\gamma$. Starting with a Calabi-Yau background with
two isometries and no $B$-field, we obtain a new background with 
$\sqrt{g^\prime} = G \sqrt{g}$ and $B^\prime = \gamma g G$ where $G=1/(1+\gamma^2 g)$. The T-dual solution is then
\beq\label{metT}
ds_6^2 = G g_{ij} e_{\varphi_i} e_{\varphi_j} + d\tilde s_4^2 = G( y_1^2+y_2^2) + d\tilde s_4^2\, .
\eeq   

Formulae for the explicit action of the T-duality
group on various quantities are collected in Appendix B. 
We need in particular the action of T-duality on the supersymmetry spinors.
Using the formulae given in the Appendix, we can show that
\bea
\label{Tspinors1}
\eta^{1 \prime}_+ &=&  \eta^{1}_+ \, ,\\
 \eta^{2 \prime}_+ &=& 
   \Omega_T  \,  \eta^{2}_+ \, ,
\eea
where
\begin{equation}\label{Ome}
\Omega_T = \frac{1}{\sqrt{1+\gamma^2 g}} \left ( 1+ \gamma \Gamma_{\varphi_1\varphi_2}\right ) \, .
\end{equation}

If $\eta^0_+$ is the $U(1) \times U(1)$ invariant spinor of the original CY, after
T-duality  we have
\begin{equation}
\eta^1_+=\eta^0_+ \, ,\qquad\qquad
\eta^2_+=-i\frac{1}{\sqrt{1+\gamma^2 g}} \left (\eta^0_+ +\gamma \sqrt{g} \frac{z}{2} \eta^0_-\right ) \, ,
\label{MLspinor}
\end{equation}
where, similarly to Section \ref{relevant}, 
we have rewritten the action of the gamma-matrices in terms of  
a normalized vector $z$
\begin{equation} \label{vect}
\Gamma_{\varphi_1\varphi_2}\eta^0_+ = \sqrt{g} \, \Gamma_{y_1y_2}\eta^0_+ 
= \sqrt{g} \, \frac{z}{2} \, \eta^0_- \, .
\end{equation}

The spinors (\ref{MLspinor}) satisfy condition (\ref{condition}) and thus  
can be brought to the form (\ref{spinoransatz2})
\begin{equation} \label{t-spin}
\eta^1_+ =a \eta_+  +b \frac{z}{2} \eta_- \, ,
\qquad\qquad \eta^2_+ =-i (a \eta_+ - b \frac{z}{2} \eta_-) \, ,
\end{equation}
with 
\begin{eqnarray}\label{rot}
\eta_+  &=&\frac{  \bar a \eta^0_+ - b (z/2) \eta^0_-}{|\eta_0^2|} \, ,\nonumber \\
b &=& \frac{-\gamma \sqrt{g}}{1+ \sqrt{1+\gamma^2 g}} \bar a \, ,\nonumber\\
|a|^2 &=& \frac{1+\sqrt{1+\gamma^2 g}}{2\sqrt{1+\gamma^2 g}}|\eta^0_+|^2 \, ,
\end{eqnarray}
where $|\eta^0_+|^2=e^{A}$. This transformation is just a rotation
in the U(1) family of almost complex structures allowed by the SU(2) structure.
By comparison with the ansatz (\ref{spchoice}), we  can also compute
\bea\label{sincos}
\sin 2 \phi &=& -\frac{\gamma\sqrt{g}}{\sqrt{1+\gamma^2 g}} \, ,\nonumber\\ 
\cos 2 \phi &=& \frac{1}{\sqrt{1+\gamma^2 g}} \, .
\eea

We want now to construct the pure spinors after the action of T-duality.  
These can be easily obtained from (\ref{puresp}) by tensoring the transformed spinors (\ref{MLspinor}).
It is however possible to compute directly the action of T-duality on the pure spinors using the results of \cite{hassan}.
This is because, via Clifford map, pure spinors can be thought as bispinors. Specializing the action of T-duality on 
bispinors as given in \cite{hassan} we have
\beq
\label{purespt}
{\rm T:} \,\,\,\,\,\,\,\,\,  \Phi_{\pm} \rightarrow  \Phi_{\pm} \Omega^{\dagger}_T \, ,
\eeq
where $\Omega_T$ is given in (\ref{Ome}) and it acts on the pure spinors by Clifford multiplication 
from the right.

As the T-duality was applied to flat space (or more generally to a CY) we should start from the standard pair of pure spinors
$(\Phi_+, \Phi_-)$ corresponding to SU(3) structure, namely the exponentiated fundamental form 
and the holomorphic three-form. In our notations, these are obtained (modulo
normalization) by taking $a=1,x=- i$ and $b=y=0$.
In a background with two isometries, if we write the metric in diagonal form
\beq
\label{dm}
{\rm d}s_6^2 = x_1^2 + x_2^2 +y_1^2 + y_2^2 + z \bar{z} \, ,
\eeq
where $z$ is the normalized vector introduced in (\ref{vect}) and $y_i$ correspond to the fiber directions along 
which T-duality acts, $\Phi_{\pm}$ read 
\bea
\label{pippo}
&& \Phi_+ =  \frac{i}{8} e^{-iJ}= \frac{i}{8} e^{z {\bar z}/2} e^{-ij} \, ,\nn\\
&& \Phi_-=  - \frac{1}{8} z\wedge \omega \, ,
\eea
with \footnote{In determining the phase in $\omega$ we used definition (\ref{formsSU2})
and the fact that $\chi_0= z \eta_0^*/2=\Gamma_{y_1y_2}\eta_0^*$.}
\bea
\label{pippo2}
&& j = x_1 \wedge y_1 + x_2 \wedge y_2 \, ,\nonumber\\ 
&& \omega = i ( x_1 + i y_1 ) \wedge ( x_2 +i  y_2 ) \, .
\eea
Note that, even though the physical context is different, the action of the second piece in $\Omega_T$, which essentially 
amounts to taking the Hodge star along the fibers,  is exactly the same as for the maximally type-changing T-duality 
action discussed in  \cite{fmt}, which exchanges  $ e^{-ij}$ and $\omega$. 
Indeed, on the forms written above it is not hard to verify that 
$ (e^{-ij})\Gamma_{y_1 y_2}^\dagger= -i\omega$ 
and $ ( \omega) \Gamma_{y_1 y_2}^\dagger = -i e^{-ij}$, so that we obtain
\footnote{ We can understand the general action of $\Gamma_{y_1 y_2}$ on $e^{-ij}$ 
and $\omega$ without referring to 
a particular basis as done above. We may use the fact that these forms 
are self-dual when restricted to the four dimensions 
transverse to the directions spanned by the complex vector $z$. Moreover, their respective real and imaginary parts form 
a full basis of self-dual forms: $(1-j^2/2)$, $j$, ${\rm Re} \, \omega$ and ${\rm Im} \, \omega$.
The star in the fiber directions only
(i.e. the action of $\Gamma_{12}$) is  a maps among self-dual forms 
but mixes forms of different rank.
So the transformed pure spinor will simply be a combination of $(1-j^2/2)$, $j$, ${\rm Re} \, \omega$ 
and ${\rm Im} \, \omega$ as in (\ref{puresp}).}
\bea\label{ppp}
&&   \Phi_+\, \Omega_T^\dagger =  \frac{i}{8\sqrt{1+\gamma^2 g}}  
e^{z {\bar z}/2}\Big[e^{-ij} 
- i\gamma \sqrt{g}  \omega \Big] \, ,\nonumber\\ 
&& \Phi_-\,  \Omega_T^\dagger =\frac{- 1}{8\sqrt{1+\gamma^2 g}}  z\wedge \Big[ \omega -i \gamma \sqrt{g} e^{-ij} \Big] \, .
\eea

Finally we can bring the pure spinors to the general form that has been used throughout  the paper  by
making use of the connection between the rotation of the four-dimensional complex structure  and the $\gamma$ parameter given in  (\ref{rot}) and (\ref{sincos}). 
This amount to a redefinition of the  real two-forms
\begin{equation}\label{rotation}
\left(\begin{array}{cc}
 \cos 2 \phi & -\sin 2 \phi \\
\sin 2 \phi &  \cos 2 \phi 
\end{array}\right) \left(\begin{array}{c}
 j\\
{\rm Re}\, \omega 
\end{array}\right) \longrightarrow \left(\begin{array}{c}
 j\\
{\rm Re} \, \omega 
\end{array}\right) \, .
\end{equation}
The phases $\alpha$ and $\beta$ are set to zero here, so we are dropping the hats on 
$\omega$.

In principle we could have just stopped this discussion here, but before 
proceeding to the discussion  of the SU(2) structure solution,
we would like to present an interpretation of the LM transformation (\ref{mlsl2}) in terms of  Generalized Complex Geometry. The idea is to describe the change of type of the pure spinors given by (\ref{ppp}) 
in terms of the standard O(2,2) action on $TM \oplus T^*M$ (where $M$ is the $T^2$ along which we 
T-dualize) \footnote{In principle GCG (or simply geometrical) descriptions 
can be problematic when a $B$-field 
with two legs along the fibers is involved, as it  is the case here. However 
since the latter is generated by SL(2) 
rotations and is constant along the fibers we will see that the geometrical description is perfectly adequate here.}. 
As already mentioned, the nontrivial part of the $\Omega_T$ action is given by the Hodge star along the fibers which 
in turns is conveniently captured by the Clifford action of $\Gamma_{y_1 y_2}$. When acting on the forms (pure spinors) 
the product of gamma-matrices gives terms with full anti-symmetrization, full contraction and partial contraction. 
Their three O(2,2) counterparts are \cite{Gualtieri}:

\noindent $\bullet \,\,\,\,\,$ $B$-transform -  a shear transformation on the cotangent bundle  $T^*M$ which acts by wedging the forms with an exponentiated two form and does not change the type of the pure spinors;

\noindent $\bullet \,\,\,\,\,$ $\beta$ transform - a shear transformation on the tangent bundle $TM$ which acts by  contracting  the forms with an exponentiated bivector and does change the type of the pure spinors;

\noindent $\bullet \,\,\,\,\,$ SL(2) rotation $A$ - a vector--valued one-form which acts by a contraction and a wedge.

\noindent The form of these operators is particularly simple in two dimensions:
\beq\label{BbetaA}
B = y_1 \wedge y_2,
 \qquad  
\beta= \iota_{y_1} \wedge\iota_{y_2},
 \qquad A= -(y_1 \wedge \iota_{y_2} - y_2 \wedge \iota_{y_1}) \, ,
\eeq
where $\iota$ denotes a contraction. Their actions on the pure spinors are 
as follows (we ignore the $z$ contribution which is inert under these actions):
\beq\label{BbetaAeij}
B (e^{-ij})= y_1 \wedge y_2 \, ,
 \qquad  
\beta (e^{-ij}) = -x_1 \wedge x_2 \, ,
 \qquad A (e^{-ij}) = i(x_2 \wedge y_1 - x_1 \wedge y_2) \, ,
\eeq
\beq\label{BbetaAomega}
B (\omega) = i x_1 \wedge x_2 \wedge y_1 \wedge y_2 \, ,
\qquad  \beta (\omega) = i \, ,
\qquad A (\omega) = x_1 \wedge y_1 + x_2 \wedge y_2 \, .
\eeq
One can define the combined action of these operations on the pure spinors 
(\ref{pippo}) as
$$
 e^{\hat B} \Phi_{\pm} := e^{2 \phi (B+\beta + A)} \Phi_{\pm}  \, .
$$
Using  that $A(B+\beta) = (B+\beta)A = 0 $ on all the 
forms and $(B+\beta +A)^2 \Phi_{\pm} = -  \Phi_{\pm}$, we find 
\beq\label{decomp}
e^{\hat B} \Phi_{\pm} = [\cos 2 \phi  
+ \sin 2 \phi (B + \beta +A)] \Phi_{\pm} \, .
\eeq
Using the explicit actions (\ref{BbetaAeij}) and (\ref{BbetaAomega}) on 
$\Phi_+$ and $\Phi_-$, we recover the 
form  (\ref{ppp}) for the transformed pure spinors provided that we identify 
the parameter $\phi$ as in (\ref{sincos}). 
Note that $\Omega_T$ does not break into series of $B$-transforms, 
$\beta$-transforms (these two do not commute!) and SL(2) rotations. 
Instead its action is given by a Clifford multiplication on pure spinors which can be seen 
as a type-changing  ``generalized $\hat B$-transform".  While the decomposition  
 (\ref{decomp}) was proved using a specific basis for $j$ and $\omega$, it is 
expected to hold more generally. It would be curious to see if a similar 
decomposition may hold for (at least a class of) higher-dimensional T-dualities.

 \subsection{The geometrical structure of the LM solution}
 
In this Section we analyze the geometry of the LM solution.
We use the rescaled
six-dimensional metric $\mbox{d} s_6^2 = e^{-2 A} \mbox{d} \tilde{s}_6^2$.

The original Calabi-Yau cone for ${\cal N}=4$ SYM is just $\mathbb{C}^3$.
We choose three complex coordinates $z_i=r \mu_i e^{i\phi_i}$ representing the three adjoint scalar fields $\Phi_i$.  A convenient
parametrization for the $z_i$ is
\bea\label{compl}
z_1&=&r \cos\alpha e^{i(\psi-\varphi_2)}\nonumber \, ,\\ 
z_2&=&r \sin\alpha\cos\theta  e^{i(\psi+\varphi_1+\varphi_2)} \, ,\nn \\ 
z_3&=&r \sin\alpha\sin\theta  e^{i(\psi-\varphi_1)} \, .
\eea
In this parametrization, the two $U(1)$ symmetries  act by shifting
$\varphi_1$ and $\varphi_2$, respectively. 

The form of the vector $z$ can be determined by equation (\ref{vect}).
By expressing the action of $\Gamma_{\varphi_1\varphi_2}$ in the
complex basis given above we find 
\beq 
z\sim dZ\, , \qquad\qquad\qquad  
Z = z_1 z_2 z_3 = \mu_1 \mu_2 \mu_3 e^{3 i \psi} \, .
\eeq

Notice that the 
complex structure on $\mathbb{C}^3$ given by the coordinates $z_i$ is the one determined by
the spinor  $\eta^0_+$.
By definition, this means that it is the Clifford vacuum 
annihilated by the complexified gamma-matrices associated to $z_i$. A simple
computation using the explicit parametrization (\ref{compl}) then gives
the result quoted above.
We can then adapt our metric to the four plus two structure determined by the
vector $z$ by considering the following vielbeins
\bea
\chi_1 &=& e^{-i\phi_1} \sqrt{\frac{g}{\mu_1^2 (\mu_2^2+\mu_3^2)}}  
(dz_1 -\frac{\bar{z}_2 \bar{z}_3dZ}{r^4 g}) \equiv x_1+ i y_1 \, ,\nonumber\\
\chi_2 &=&  e^{-i \phi_2}\sqrt{1+\frac{\mu_3^2}{\mu_2^2}}
(dz_2-\frac{\bar{z}_1\bar{z}_3 dZ}{r^4 g}) 
+ \frac{\mu_3^3  e^{-i\phi_1}}{\mu_1 \sqrt{\mu_2^2+\mu_3^2}} 
(dz_1 -\frac{\bar{z}_2\bar{z}_3dZ}{r^4 g}) \equiv x_2 + i y_2 \, ,\nonumber
\eea
where $$g=\mu_1^2 \mu_2^2 +\mu_2^2 \mu_3^3 +\mu_3^2 \mu_1^2 = \sin^2\alpha (\cos^2\alpha +\sin^2\alpha \sin^2\theta\cos^2\theta) \, .$$ 
These combine
with $z$ to give an integrable complex structure on $\mathbb{C}^3$
\beq\label{diago}
ds_6^2 =\sum_{i=1}^3  dz_i d\bar{z}_i = 
\chi_1\bar{\chi}_1 + \chi_2 \bar{\chi}_2+\frac{ dZ d\bar{Z}}{r^4 g} \equiv x_1^2 + x_2^2 + y_1^2 + y_2^2 +\frac{ dZ d\bar{Z}}{r^4 g} \, .
\eeq
The metric on $T^2$ is simply given by the terms $y_1^2 + y_2^2$.
Explicitly
\bea
y_1 &=& r \sqrt{\cos^2\alpha+\sin^2\alpha \sin^2\theta \cos^2\theta} 
(d\varphi_2 -\frac{\cos^2\alpha -2 \sin^2\alpha \sin^2\theta \cos^2\theta}{\cos^2\alpha + \sin^2\alpha \cos^2\theta \sin^2\theta}d\psi) \, ,\nonumber\\
y_2 &=& r \sin\alpha (d\varphi_1+\cos^2\theta d\varphi_2 
+\cos 2\theta d\psi) \, .\nonumber
\eea
We now perform the T-duality transformation (\ref{mlsl2}) to the 10 dimensional
metric $$ds^2= r^2 ds_4^2 + \frac{1}{r^2} ds_6^2\, .$$
The original $B$-field is zero and the $\nu$ parameter of the 
two-torus is simply given by $\nu = i \sqrt{g}$. 
The six-dimensional internal metric after T-duality
$$ds^2= r^2 ds_4^2 + \frac{1}{r^2} ds_{ML}^2$$
with its natural almost complex structure is now given by
\bea \label{csML}
&& ds_{LM}^2 = \chi_1^\prime\bar{\chi}_1^\prime+ \chi_2^\prime\bar{\chi}_2^\prime + \frac{ dZ d\bar{Z}}{r^4 g} \equiv x_1^2 +x_2^2 + G(y_1^2 +y_2^2) + \frac{ dZ d\bar{Z}}{r^4 g} \, ,\nonumber\\
&& \chi_i^\prime = x_1 +i \sqrt{G} \, y_i \, .
\eea
In real coordinates we can also write the metric as in \cite{LM}
\beq 
\label{eqML}
ds_{LM}^2 = dr^2 +r^2 ( \sum_{i=1}^3 (d\mu_i^2 + g \mu_i^2 d\phi_i^2) + 9 \gamma^2 G \mu_1^2\mu_2^2\mu_3^2 d\psi^2 ) \, .\eeq
The expressions for the other fields can be easily found by using the rules for T-duality given in Appendix B and read
\bea\label{other}
e^{2 A} &=& r^2 \, ,\nonumber\\
e^{\phi} &=& \sqrt{G} \, ,\nonumber\\
B_2 &=& \gamma\sqrt{g} G \frac{y_1\wedge y_2}{r^2} \, ,\nonumber\\
F_3 &=& 12 \gamma \cos\alpha \sin^3\alpha \sin\theta\cos\theta 
d\psi\wedge d\alpha \wedge d\theta \, ,\nonumber\\
F_5 &=& 4(\rm{vol} AdS_5 + * \rm{vol} AdS_5) \, .
\eea
\noindent It is immediate to check that the vector 
\beq z= \frac{dZ}{r^2 \sqrt{g}} \eeq
is conformally closed (equation (\ref{oneform}))
\beq   
 \mbox{d}\Big(e^{2 A-\varphi} e^{i (\alpha + \beta)} \sin 2 \phi z \Big) =0 
\eeq
with $\alpha=\beta=0$. From equation (\ref{supo}) we also find
\beq 
{\mbox d} W  \sim e^{2 A-\varphi} e^{i (\alpha + \beta)} \sin 2 \phi  z = {\mbox d} ( -\gamma z_1 z_2 z_3)
\eeq
 which exactly corresponds to the (abelianized) superpotential
$W\sim \Phi_1\Phi_2\Phi_3$ in  (\ref{beta}). 

The metric is not K\"ahler but the condition of supersymmetry
(\ref{twoform}) requires the existence of a conformally closed two form
 \beq  
\mbox{d} \Big( e^{-\varphi} \hat J \Big) =  \mbox{d} \Big[ e^{-\varphi}
 \left ( j +\frac{i}{2} \cos2 \phi z \wedge \bar{z}\right ) \Big]=0  \, .\eeq
This equation is also trivially satisfied since
\beq 
\hat J = \frac{i}{2} (\chi_1^\prime\bar{\chi}_1^\prime + \chi_2^\prime\bar{\chi}_2^\prime + \sqrt{G} z\bar{z}) = \sqrt{G} (x_1 y_1 +x_2 y_2 + \frac{i}{2} z\bar{z}) = e^{\varphi} J_{N=4} \, . \eeq

It is then straightforward to check that all other conditions for supersymmetry
(\ref{ffive}-\ref{Ifourform}) are satisfied with the definition (\ref{pippo2}).  For this check we used the almost complex structure defined in equation (\ref{csML}). As discussed in details in the previous Section, our supersymmetry conditions are satisfied with a specific choice of almost complex structure,
which is  obtained by applying a rotation (\ref{rot}-\ref{rotation}) to  the one defined by the spinor $\eta_0$. An explicit computation shows that the result is indeed  the almost complex structure defined in formula (\ref{csML}).

We can also analyze the moduli space for probe D3 branes in this 
background. The general formula (\ref{moduli}) requires
\beq \sin 2 \phi = -\gamma G \sqrt{g} \equiv 0 \, ,\eeq
or equivalently
\beq g = \mu_1^2 \mu_2^2 +\mu_2^2 \mu_3^2 +\mu_3^2 \mu_1^2 \equiv 0 \, , \eeq 
which  determines the three branches
\beq z_i=z_j=0, \, z_k\ne 0\, , \qquad \qquad i\ne j\ne k \eeq
in agreement with the field theory analysis.

As already mentioned, for rational $\gamma$ 
there are other Coulomb branch vacua. These have been identified in \cite{LM} as $n_5$ D5 branes wrapped on the two torus (in points where it is not vanishing)  
with $1/\gamma$ units of gauge flux. These objects carry both units of D5 charge
($n_5$) and units of D3  charge ($n_5/\gamma$) and satisfy the charge 
quantization condition only for $\gamma$ rational.
We can easily check that these are BPS states using conditions (\ref{Dp}).
For a D5 brane wrapped on the two torus and the spinorial ansatz (\ref{spinoransatz2}), the supersymmetry conditions become
$  {\rm Re} \hat J =0 $
which is automatically
satisfied and
\beq
\sin 2\phi ({\it F}-B) +\cos 2\phi G y_1\wedge y_2 =0
\eeq
which, given the form of the B field (\ref{other}), is satisfied
exactly by ${\it F}= y_1\wedge y_2/(\gamma \sqrt{g})$.  
 
A marginal deformation analogous to the $\beta$-deformation 
exists for all quiver gauge theories associated with toric Calabi-Yau
singularities. All these backgrounds have indeed two isometries commuting 
with the supersymmetry generators, or equivalently, from the field theory point
of view, two U(1) global symmetries in addition to the R-symmetry. The
generating technique of \cite{LM} then determines the supergravity dual
of the marginally deformed conformal field theory. The above analysis of
the geometrical structure of the supersymmetric solution will apply to all
these backgrounds with minor changes. 

\section{Conclusions}

In this paper we started a detailed analysis of the conditions of 
supersymmetry for SU(2) structure backgrounds. While our results are not
the most general ones, since the solution relies on a particular choice in the spinorial ansatz (\ref{spinoransatz2}), we were able to write a very simple set of equations
for the SU(2) invariant forms and the fluxes for a very large class of
backgrounds. In particular, generalizing results in \cite{GW}, we described
a simple class of complex manifolds and associated fluxes which solve the
supersymmetry conditions of type IIB and are characterized by the existence of
a generalized K\"ahler potential. 
 
The use of G-structures has been
useful in the past not only for characterizing known solutions but also 
for finding new ones. One example is the baryonic 
branch of the Klebanov-Strassler solution \cite{bgmpz}.  
In this paper, we studied the geometry of the PW and LM solutions describing massive and marginal deformations of conformal theories. 
It would be quite interesting to
pursue further the analysis of this paper and try to find new conformal
and non conformal backgrounds. There are various obvious directions 
where one could move. For example, there should be a large number of conformal
gauge theories corresponding to warped $AdS_5$ solutions with fluxes. 
 We
gave a possible characterization of those solutions corresponding to massive
deformations of conformal gauge theories in Section 4. PW is the only known solution and it would be interesting
to find new examples.
Moreover, there exist other marginal deformations of ${\cal N}=4$ SYM and other
conformal theories beside the $\beta$-deformation. In particular, it would be
interesting to find the marginal deformation of ${\cal N}=4$ 
associated with the coupling $h^\prime$ in equation (\ref{beta2}).

We are only making the first steps understanding the interplay between the GCY geometry 
of the internal manifolds and that of moduli spaces of the  dual gauge theories.
Much of the information about the solution, including the type of the deformation in the gauge theory, is encoded in the pure spinors on the internal six-dimensional manifold.  Moreover for the case of the $\beta$-deformations,  the solution generating T-duality transformation can be seen as a type-changing generalized $\hat B$ transform that acts on both the pure spinors and the RR fields.
The use of Generalized Complex Geometry  is also a necessary
ingredient in studying supersymmetric branes and calibration conditions in flux backgrounds  \cite{martucci}.    This should be particularly important in solutions with interesting topologies where
BPS states can correspond to branes wrapping non-trivial supersymmetric 
 cycles. For example, using wrapped D3 branes one can determine
the exact dimensions of fields in duals of conformal field theories. 
From this point of view, the simplicity of solutions in Section 4, 
which are characterized by a generalized K\"ahler 
potential, suggests that it is perhaps possible to make a general analysis
of volumes and R-charges analogous to that in \cite{MSY}. Moreover,
it should be also quite  straightforward to verify, using the generalized calibrations provided by the pure spinors, that central and R-charges do not change for marginal deformations.

Finally, this paper is restricted to the study of conformal case, but our
equations and formalism have obvious applications to the non conformal case
as well. 

\vskip 0.5cm
\noindent 
{\bf Acknowledgments}
\vskip 0.3cm

\noindent We would like to thank G. Dall'Agata, P. Grange, L. Martucci, 
R. Russo and 
A. Tomasiello
for helpful discussions. A.Z. is supported in part by INFN and MURST under 
contract 2005-024045-004. A.Z thanks the GGI Center of Physics
in Florence where part of this work was done.
R.M. and A.Z. are supported in part by 
the European Community's Human Potential Programme
MRTN-CT-2004-005104, and M.P. by MRTN-CT-2004-512194.

\appendix
\section*{Appendices}

\section{The general conditions of supersymmetry}
\label{app:susy}

In this short Appendix we report the condition of supersymmetry corresponding
to the most general SU(2) structure ansatz for the spinors 
(\ref{spinoransatz}). From equations (\ref{purespeq}) we obtain the following
set of differential constraints for the SU(2) invariant forms and the fluxes.

The equation for $\Phi_-$ gives 
\bea
\label{2form}
&& \mbox{d}\Big(e^{2 A - \varphi} ( b x - a y ) z  \Big) =0 \\
\label{4form}
&& 
\mbox{d}\Big[ e^{2 A - \varphi} ( b y \bar{\omega} - a x \omega + ( b x + a y ) j) z \Big]
+ i  ( b x - a y ) H z = 0 \\
\label{6form}
&& \mbox{d}\Big[  e^{2 A - \varphi}  ( b x - a y ) z j^2 \Big] 
- 2 i  H z \Big[  b y \bar{\omega} - a x \omega + ( b x + a y ) j \Big] = 0
\eea
and that for $\Phi_+$
\bea
\label{1form}
&& e^{-2 A + \varphi} \mbox{d} \Big[e^{2A - \varphi} (a \bar{x} + b \bar{y})\Big] = \mbox{d} A (x \bar{a} + y \bar{b}) + \frac{e^{\varphi}}{2} \Big[ a_- F_1
-i  a_+ *F_5 \Big] \\
\label{3form}
&& e^{-2 A + \varphi} \mbox{d} \Big[e^{2A - \varphi} 
( (a \bar{x} - b \bar{y}) j + \frac{i}{2} (a \bar{x} + b \bar{y}) z \bar{z} 
+ a \bar{y} \omega + b \bar{x} \bar{\omega}) \Big] \\
&& \quad \quad 
- i  (a \bar{x} + b \bar{y}) H = \nn \\
&& \quad \quad -  \mbox{d} A \Big[ (x \bar{a} - y \bar{b}) j 
+ \frac{i}{2} (x \bar{a} + y \bar{b}) z \bar{z} 
+ y \bar{a} \bar{\omega} + x \bar{b} \omega \Big] 
+  \frac{e^{\varphi}}{2} \Big[i a_- F_3
+ a_+  *F_3 \Big] \nn \\
\label{5form} 
&& e^{-2 A + \varphi} \mbox{d} \Big[e^{2A - \varphi} ( (a \bar{x} + b \bar{y}) j^2
+ i (a \bar{x} - b \bar{y}) j z \bar{z} 
+ i ( a \bar{y} \omega + b \bar{x} \bar{\omega} ) z \bar{z} \Big] \\
&& \quad \quad - 2 i  H \Big[ a \bar{y} \omega + b \bar{x} \bar{\omega} 
+ (a \bar{x} - b \bar{y}) j 
+ \frac{i}{2} (a \bar{x} + b \bar{y}) z \bar{z} \Big] = \nn \\
&& \quad \quad  \mbox{d} A \Big[ (x \bar{a} + y \bar{b}) j^2 
+ i  (x \bar{a} - y \bar{b}) j z \bar{z} 
+ i (y \bar{a} \bar{\omega} + x \bar{b} \omega ) z \bar{z} \Big] 
- e^{\varphi} \Big[ a_- F_5 - i a_+ *F_1 \Big] \nn
\eea
with $a_{\pm} = |a|^2 +|b|^2 \pm (|x|^2 +|y|^2 )$

Note that the equations of motion for the RR fluxes follow from the pure spinor equations (\ref{purespeq}).
However in order to find a complete solution, the NS-flux equation of motion  
and the Bianchi identities for the fluxes,
$dH=0$ and $(d-H)F=0$, must still be imposed.

\section{Formulae for the T-duality}

In this Appendix we
collect formulae for the explicit action of the T-duality
group on various quantities using \cite{hassan}. 
The T-duality group of $T^2$ is
O(2,2)$\equiv$ SL(2,R)$\times$ SL(2,R), which can be represented as the set of 
matrices O such that $O^T J O=J$ where
\begin{equation}
J=\left(\begin{array}{cc}
O & I_{2\times 2} \\
I_{2\times 2} & O
\end{array}\right) \, .
\end{equation}
A matrix with two by two blocks
\begin{equation}
O=\left(\begin{array}{cc}
A & B \\
C & D
\end{array}\right)
\end{equation}
 acts on the two by two matrix
$E=g+B$ as \cite{Treport}
\beq \label{Tmet}
E\rightarrow (A E+B) (C E +D)^{-1} \, .
\eeq 
The SL(2,R) subgroup that acts on the complexified
K\"ahler modulus of the torus can be parametrized as \footnote{It is well
known that the second copy of SL(2,R) acts on the complex structure of the 
two torus
\begin{equation}
M=\left(\begin{array}{cc}
\alpha & \beta \\
\gamma & \delta
\end{array}\right),\qquad\qquad 
O=\left(\begin{array}{cc}
M & 0 \\
0 & M^{T\, -1}
\end{array}\right), \qquad\qquad E\rightarrow M E M^T
\end{equation}
 by transforming $g\rightarrow M g M^T$. This is just a change 
of coordinates and thus less useful for generating new solutions.}
\begin{equation}
  O=\left(\begin{array}{cc}
a I & b \epsilon \\
-c \epsilon  & d I
\end{array}\right),
\end{equation}
where $\epsilon$ is the Pauli matrix $i\sigma_2$ and $ad-bc=1$. 

It is convenient to write 
\begin{equation}
O_{LM} = T^{-a} O T^{-a}
\end{equation}
where
\begin{equation}
O=\left(\begin{array}{cc}
(S+R)/2 & (S-R)/2\\
(S-R)/2   & (S+R)/2
\end{array}\right),
\label{hassan}
\end{equation}
with $\gamma=2 a/(1+a^2)$ and
\begin{equation} S+R=2\sqrt{1-\gamma^2} I,\qquad S-R=-2 \gamma 
\epsilon \end{equation} and $T$ is the standard shift generator T of SL(2;R)
$$
T^{-a} = \left(\begin{array}{cc}
I_{2 \times 2} & -a \epsilon\\
0  & I_{2 \times 2}
\end{array}\right) \, .
$$
The advantage of this parametrization is that for matrices of the form (\ref{hassan}) the T-duality transformations considerably simplify. The part
of the transformation corresponding to $O$ is the only one that acts non 
trivially on the coordinates. $T^{-a}$ just shifts the B field and can be
easily superimposed once the action of $O$ is known. 

Consider now $S,R$ and $O$ as six-dimensional matrices by trivial extension  
${\cal R}= (I_{4\times 4},R)$
and ${\cal S}= (I_{4\times 4},S)$. Defining the six dimensional matrix 
\cite{hassan}
\beq\label{Q} Q =  \frac{1}{2} [({\cal S} +{\cal R}) +({\cal S} -{\cal R})(g_6 + B_6)] \, ,\eeq
and the operator
\bea\label{Omeg}
&& \Omega_T = \frac{1}{2} \sqrt{ \frac{\det B}{\det Q }} \left ( 1 - \frac{1}{2} A^{ij} \Gamma_{ij}\right ) \, ,\nonumber\\
&& B^{ij} = [({\cal R}+{\cal S})+({\cal R}-{\cal S})B_6)]_{ij} \, ,\nonumber\\
&& A^{ij} = [({\cal S}-{\cal R})^{-1}({\cal S} +{\cal R}) +B_6]^{-1}_{ij} \, ,
\eea
the effect of T-duality on the other fields and the supersymmetry parameters can be summarized by \cite{hassan}
\bea
\label{Tmetrica}
g_6^\prime &=& Q^{T -1} g_6 Q^{-1} \, ,\\
\label{Tdilaton}
e^{\phi^\prime} &=&\frac{e^{\phi}}{\det Q} \, ,\\
\label{TRR}
F^\prime &=& \sqrt{\det Q}\, \Omega_T F\nonumber \\
\label{Tspinors}
\eta^{1 \prime}_+ &=& \eta^{1}_+ \, ,\\
 \eta^{2 \prime}_+ &=& 
     \Omega_T   \eta^{2}_+ \, .
\eea
where $F$, as usual, is the formal sum of RR fields strength, and 
we also gave an expression for the transformation of the metric alternative to (\ref{Tmet}).

Let us apply these formulae to our case. Starting
with flat space and zero $B$-field, the action of $T^{-a}$ shifts the
value of the $B$-field to $-a$. So we apply the previous formulae
for the action of  $O$ 
to a background with $-a$ $B$-field. 
By direct computation det$\, Q = 1+\gamma^2 g$ and
\begin{equation}
\Omega_T = \frac{1}{\sqrt{1+\gamma^2 g}} \left ( 1+ \gamma \Gamma_{\varphi_1\varphi_2}\right ) \, .
\end{equation}
It is also easy to check that the transformation on the metric reproduces
equation (\ref{eqML}).

\end{document}